\documentclass[a4page, 10pt]{article}
\usepackage{amssymb,amsmath}
\usepackage{graphicx}
\numberwithin{equation}{section}

\begin{document}

\title{A Possible Quantum Effect of Gravitation}

\author{Jarmo M\"akel\"a\footnote{Vaasa University of Applied Sciences,
Wolffintie 30, 65200 Vaasa, Finland, email: jarmo.makela@vamk.fi}}

\maketitle

\begin{abstract}

Beginning from the standard Arnowitt-Deser-Misner (ADM) formulation of general relativity we construct a tentative model of quantum gravity from the point of view of an observer with constant proper acceleration, just outside of a horizon of spacetime. In addition of producing the standard results of black-hole thermodynamics, our model makes an entirely new prediction that there is a certain upper bound for the energies of massive particles. For protons, for instance, this upper bound is around $1.1\times 10^{21}eV$. The result is interesting, because this energy is roughly of the same order of magnitude as are the highest energies ever measured for protons in cosmic rays. 
 
\end{abstract}

\maketitle

\section{Introduction}

The most important aim of any theory of physics is to make predictions, which could be proved either right or wrong by means of experiments and observations. So far it has always been thought that all quantum effects of gravity lie in the Planck regime, where the typical energies are of the order of $10^{28}eV$, and therefore well beyond any possibilities for observations and experiments. In this paper we speculate on a possiblity that there is at least one effect of quantum gravity, where the typical energies are well below the Planck energy, and measurable by means of the current observational techniques.

To be more precise, we consider an idea that quantum mechanics, when applied to Einstein's general theory of relativity, could imply that there is a certain upper bound for the velocities of massive particles, which is a bit {\it less} than the speed of light.  As a consequence, any massive particle with mass $m$ has a certain {\it maximum energy}. We suggest that to make the predictions of our model of quantum gravity consistent with the well-known results of black-hole thermodynamics, this maximum energy must be:
\begin{equation}
E_{max} = mc^2\cosh \left[\frac{2\pi^2}{\ln(2)}\right].
\end{equation}
For protons, for example, this maximum  energy is around $1.1\times 10^{21}eV$, and therefore seven orders of magnitude below the Planck energy. The result is interesting, because the highest energies ever measured for protons in cosmic rays are roughly of the same order of magnitude, bur still less than the upper bound set by Eq. (1.1): Now and then energies of order $10^{19}eV$ have been measured for protons, \cite{yy} and the highest energy ever measured has been $(3.2\pm 0.9)\times 10^{20}eV$. \cite{kaa} Thus the prediction given by Eq. (1.1) could indeed be proved wrong by means of technologies available today. Should one ever observe a proton with energy higher than $1.1\times 10^{21}eV$, the model of quantum gravity leading to Eq. (1.1) would have been proved erroneous.

    The derivation of Eq. (1.1) in this paper is based on a Hamiltonian approach to quantum gravity. We construct the Hamiltonian operator for the gravitational field from the point of view of an observer just outside of a horizon of spacetime. The key assumption of our model 
is that the proper acceleleration $a$ of the observer is constant. In other words, no matter what may happen in spacetime, the proper acceleration $a$ of the observer is always the same. Our model of quantum gravity involves a free, dimensionless parameter, which must be fixed such that the model reproduces the familiar results of black-hole thermodynamics. It turns out that this parameter determines an upper bound for the velocities of massive particles with respect to the observer.  It must be emphasized that even though our results may seem startling, they have nevertheless been obtained by means of a straightforward application of the standard rules of quantum mechanics to Einstein's general relativity.

Unless otherwise stated, we shall always use the natural units, where $\hbar=G=c=k_B=1$.

\section{Hamiltonian}

The first successfull Hamiltonian formulation of Einstein's general relativity was constructed by Arnowitt, Deser and Misner in the year 1962. \cite{koo} The starting point of this  formulation is to write the line element of spacetime as:
\begin{equation}
ds^2 = -N^2\,dt^2 + q_{ab}(dx^a + N^a\,dt)(dx^b + N^b\,dt).
\end{equation}
In this line element, $q_{ab}$ $(a,b=1,2,3)$ is the metric tensor induced on the spacelike hypersurface of spacetime, where the time coordinate $t=constant$. $N$ is the lapse function, and $N^a$ is the shift vector. As it is well known, the action in this so-called ADM formulation may be cast into a Hamiltonian form as:
\begin{equation}
S_{ADM} = \int(p^{ab}\dot{q}_{ab} - N_a\mathcal{H}^a - N\mathcal{H})\,d^4x,
\end{equation}
where we have integrated over the whole spacetime. The dot means the time derivative, $p^{ab}$ is the momentum conjugate to $q_{ab}$, and we have denoted:
\begin{subequations}
\begin{eqnarray}
\mathcal{H}^a &:=& - 2p^{as}_{\,\,\,\vert s},\\
\mathcal{H} &:=&\frac{\sqrt{q}}{16\pi}(K_{ab}K^{ab} - K^2 - \mathcal{R}),
\end{eqnarray}
\end{subequations}
where "$\vert$" means the covariant derivative on the hypersurface. $K_{ab}$ is the exterior curvature tensor on the hypersurface, $K$ is its trace, and $\mathcal{R}$ is the Riemann curvature scalar on the hypersurface.

     Varying the ADM action $S_{ADM}$ with respect to the lapse $N$ gives the Hamiltonian constraint
\begin{equation}
\mathcal{H} = 0,
\end{equation}
and if we vary the ADM action with respect to the shift vector $N^a$, we get the three diffeomorphism constraints
\begin{equation}
\mathcal{H}^a = 0.
\end{equation}
The Hamiltonian constraint states the invariance of $S_{ADM}$ in the reparametrization of the time coordinate $t$, and it is equivalent with Einstein's field equation $G_{00}=0$ in vacuum. The three differomorphism constraints, in turn, state the invariance of $S_{ADM}$ with respect to the choice of the spatial coordinate $x^a$ on the hypersurface, and they are equivalent with the equations $G_{a0}=0$ $(a=1,2,3)$.

   When the spacelike hypersurface, where the time coordinate $t=constant$ is compact, we may obtain the equations $G_{ab}=0$, which are the real, dynamical equations of general relativity in the ADM formulation, quite simply, by means of the variation of $q_{ab}$ and its conjugate momentum $p^{ab}$. However, if the spacelike hypersurface either is not compact, or the spatial region under consideration possesses a {\it boundary}, we meet with a curious problem: The curvature scalar $\mathcal{R}$ on the hypersurface has the property:
\begin{equation}
\sqrt{q}\mathcal{R} = \sqrt{q}q^{mn}(\Gamma_{as}^s\Gamma_{mn}^a - \Gamma_{an}^s\Gamma ^a_{ms}) - \lbrack q^{mn}\sqrt{q}(\delta^a_m\Gamma_{ns}^s - \Gamma^a_{mn})\rbrack_{,a}.
\end{equation} 
Since the Christoffel symbol $\Gamma_{mn}^a$ on the hypersurface involves the first spatial derivatives of $q_{ab}$, the second term on the right hand side of Eq. (2.6) involves its {\it second} derivatives. Hence it follows that when varying $S_{ADM}$, we must fix not only the metric tensor $q_{ab}$, but also its {\it derivatives} on the boundary. To overcome this problem we must compensate the second term on the right hand side of Eq. (2.6) by means of an appropriate {\it boundary term}. When the lapse $N\equiv 1$ and the shift $N^a\equiv 0$ on the boundary $B$ of the hypersurface, which means that the time coordinate of an observer at rest with respect to the spatial coordinates on the boundary agrees with his proper time, an appropriate boundary term is the time integral of the function:
\begin{equation}
-H_B := \frac{1}{16\pi}\oint_B\lbrack q^{mn}\sqrt{q}(\delta_m^a\Gamma_{ns}^s - \Gamma_{mn}^a)\rbrack\,dS_a,
\end{equation}
where we have integrated over the closed two-boundary $B$ of the hypersurface. Hence the action takes the form:
\begin{equation}
S = \int(p^{ab}\dot{q}_{ab} - N_a\mathcal{H}^a - N\mathcal{H})\,d^3x\,dt - \int H_B\,dt,
\end{equation}
and we may write the Hamiltonian of the system as:
\begin{equation}
H = \int(N_a\mathcal{H}^a + N\mathcal{H})\,d^3x + H_B,
\end{equation}
where we have integrated over that part of the hypersurface, which is bounded by the closed two-surface $B$. 

    When the Hamiltonian constraint $\mathcal{H}=0$ and the three diffeomorhism constraints $\mathcal{H}^a=0$ are satisfied, the Hamiltonian of the gravitational field inside of the closed two-surface $B$ reduces to the form:
\begin{equation}
H = H_B.
\end{equation}
Solving the Hamiltonian and the diffeomorphism constraints we may separate the gauge degrees of freedom of the gravitational field from its real, physical degrees of freedom. In general, the metric tensor $q_{ab}$ has six independent components at each spacetime point, and because there are four constraints at each point, the gravitational field has just $6-4=2$ real, physical degrees of freedom at each point of spacetime. In principle, we may attempt to solve the gauge degrees of freedom from the constraints in terms of the physical degrees of freedom, and then substitute in $H_B$. Denoting those physical degrees of freedom by $\alpha^s$, we may therefore write the Hamiltonian $H_B$ in Eq. (2.10) as:
\begin{equation}
H_B = H_B(\alpha^s).
\end{equation}

   The {\it Hamiltonian reduction} described above has really been performed in spherically symmetric, asymptotically flat vacuum spacetimes by Kuchar. \cite{nee} For asymptotically flat vacuum spacetimes $H_B$ reduces to the ADM energy of spacetime. The only spherically symmetric, asymptotically flat vacuum solution to the Hamiltonian and the diffeormorphism constraints is the Schwarzschild solution, which describes spacetime involving the Schwarzschild black hole. The ADM energy in such spacetime agrees with the Schwarzschild mass $M$ of the hole. A similar Hamiltonian reduction has also been carried out in spherically symmetric spacetime interacting with spherically symmetric electromagnetic field. \cite{vii,kuu} In general, for spacetime involving a black hole the physical degrees of freedom may be chosen to be the mass, electric charge and angular momentum of the hole. In what follows, we shall take the function $H_B(\alpha^s)$ as the real, physical Hamiltonian of the gravitational field inside of a closed, spacelike two-surface embedded into a spacelike hypersurface, where the time coordinate $t=constant$, and see, where this will take us. \footnote{For the proof that  the boundary term really gives, under certain conditions, the physical Hamiltonian of the gravitational field, see Ref. \cite{ysi}.}

We shall now obtain an expression for the Hamiltonian $H_B(\alpha^s)$ defined in Eq. (2.7), when the closed two-surface $B$ lies just outside of a horizon of spacetime.
As the starting point we shall use the properties of the {\it Rindler spacetime.} As it is well known, Rindler spacetime is just a two-dimesional, flat Minkowski spacetime equipped with Rindler coordinates $t$ and $x$. When written in terms of these coordinates, the line element takes the form: \cite{seite}
\begin{equation}
ds^2 = -x^2\, dt^2 + dx^2.
\end{equation}
The Rindler coordinates $t$ and $x$ are related to the flat Minkowski coordinates $X$ and $T$ such that:
\begin{subequations}
\begin{eqnarray}
t &:=& \tanh^{-1}\left(\frac{T}{X}\right),\\
x &:=& \sqrt{X^2 - T^2}.
\end{eqnarray}
\end{subequations}
Eq. (2.13a) implies that we may understand the Rindler time coordinate $t$ as the {\it boost angle} of an observer in the Minkowski spacetime. The spatial coordinate $x$, in turn, determines the {\it proper acceleration} of the observer: An observer with constant $x$ has constant proper acceleration
\begin{equation}
a = \frac{1}{x}.
\end{equation}
So we find that at the {\it Rindler horizon}, where $x=0$, and therefore $T= \pm X$, the proper acceleration $a$ tends to infinity. 

Almost any spacetime equipped with a horizon may be approximated, just outside of the horizon, by means of an appropriate four-dimensional generalization of the Rindler spacetime. The four-dimensional generalization used in this paper is the one with the line element \cite{kasi}
\begin{equation}
ds^2 = -y\,dt^2 + \frac{(y')^2}{4y}\,d\lambda^2 + h_{jk}\,d\chi^j\,d\chi^k.
\end{equation}
When obtaining this line element we have replaced the Rindler coordinate $x$ by a new coordinate
\begin{equation}
y := x^2,
\end{equation}
which, in turn, is assumed to be a function of a spatial coordinate $\lambda$, and parameters $\alpha^s$ $(s = 1,2,\dots, M)$ determining the geometrical properties of spacetime. In other words,
\begin{equation}
y= y(\lambda, \alpha^1, \alpha^2,\dots, \alpha^M).
\end{equation}
For the Reissner-Nordstr\"om spacetime, for instance, which is an asymptotically flat, electrovacuum spacetime involving the Reissner-Nordstr\"om black hole, the parameters may be taken to be the mass and the electric charge of the hole. The coordinates $\chi^j$ $(j=1, 2)$, in turn, are the coordinates on the spacelike two-surface, where $\lambda=constant$. $h_{jk}$ $(j, k=1, 2)$ is the metric tensor induced on this two-surface, and we have assumed that $h_{jk}$ is a function of the spatial coordinates $\lambda$ and $\chi^j$, but not of the parameters $\alpha^s$. In Eq. (2.15) we have denoted:
\begin{equation}
y' := \frac{\partial y}{\partial \lambda}.
\end{equation}

  In general, the proper acceleration vector field of an observer is
 \begin{equation}
a^\mu = u^\sigma u^\mu_{;\sigma},
\end{equation}
where $u^\mu$  is the future pointing unit tangent vector field of the world line of the observer. For an observer at rest with respect to the coordinates $\lambda$ and $\chi^j$ the only non-zero component of $u^\mu$ is $u^t=1/\sqrt{y}$, and therefore the only non-zero component of $a^\mu$ for such observer is:
\begin{equation}
a^\lambda = \frac{2}{y'}.
\end{equation}
Hence the norm of $a^\mu$ is:
\begin{equation}
a = \sqrt{a_\lambda a^\lambda} = \frac{1}{\sqrt{y}}.
\end{equation}
We shall assume that the proper acceleration $a$ of our observer is always a constant, no matter what may happen in spacetime. According to Eq. (2.21) constant $a$ means constant $y$, and hence it follows that when the physical quantities $\alpha^s$ take on infinitesimal changes $d\alpha^s$, the coordinate $\lambda$ must take on a change $d\lambda$ such that
\begin{equation}
dy = \frac{\partial y}{\partial\lambda}\,d\lambda + \frac{\partial y}{\partial\alpha^s}\,d\alpha^s =0.
\end{equation}

Let us now obtain an expression for $H(\alpha^s)$ just outside of the horizon, beginning from Eq. (2.15). Using Eq. (2.7) we find:
\begin{equation}
\begin{split}
H_B &= -\frac{1}{16\pi}\oint_B\left(\frac{2\sqrt{y}}{y'}\Gamma_{\lambda j}^j - \frac{y'}{2\sqrt{y}}h^{jk}\Gamma^\lambda_{jk}\right)\sqrt{h}\,d^2\chi\\
      &= -\frac{1}{8\pi}\oint_B\frac{\sqrt{y}}{y'}h^{jk}{h'}_{jk}\sqrt{h}\,d^2\chi,
 \end{split}
\end{equation}
where we have integrated the coordinates $\chi^1$ and $\chi^2$ over the two-surface $B$. It is interesting to observe, what happens to $H_B$, when the quantities $\alpha^s$ take on infinitesimal changes $d\alpha^s$. The resulting change in the Hamiltonian $H_B$ is:
\begin{equation}
\begin{split}
dH_B &= \frac{\partial H_B}{\partial \alpha^s}\,d\alpha^s\\
         &= -\frac{1}{8\pi}\oint_B\left\lbrack\frac{1}{2\sqrt{y}y'}\frac{\partial y}{\partial\alpha^s} - \frac{\sqrt{y}}{(y')^2}\frac{\partial y'}{\partial\alpha^s}\right\rbrack h^{jk}{h'}_{jk}\sqrt{h}\,d^2\chi\,d\alpha^s.
\end{split}
\end{equation}
Because we assume that the proper acceleration $a=constant$ on the two-surface $B$, Eq. (2.22) implies:
\begin{equation}
\frac{\partial y}{\partial\alpha^s}\,d\alpha^s = -y'\,d\lambda.
\end{equation}
When we are just outside of the horizon, where $y=0$, and $y'\ne 0$, we may ignore the second term indiside of the brackets in Eq. (2.24). Hence we may write, in effect, 
\begin{equation}
dH_B = \frac{a}{8\pi}\,dA,
\end{equation}
where 
\begin{equation}
dA = \frac{dA}{d\lambda}\,d\lambda = \frac{1}{2}\oint_B h^{jk}{h'}_{jk}\sqrt{h}\,d\lambda
\end{equation}
is the change taken by the area 
\begin{equation}
A = \oint_B\sqrt{h}\,d^2\chi
\end{equation}
of the two-surface $B$, when the coordinate $\lambda$ takes on the change $d\lambda$. Since the proper acceleration $a$ of the observer is kept as a constant, we may thus write the gravitational Hamiltonian from the point of view of such observer as:
\begin{equation}
H = \frac{a}{8\pi}A,
\end{equation}
where $A$ is the area of the horizon. 

\section{Classical Dynamics}

  We shall now proceed to the study of the classical Hamiltonian dynamics of spacetime implied by Eq. (2.29). As it is well known, the so-called {\it densitized triads}
\begin{equation}
\tilde{E}^m_I := \sqrt{q}E^m_I
\end{equation}
may be used as the coordinates of the configuration space in the Hamiltonian formulation of general relativity. In that case the quantities
\[
-\frac{1}{16\pi}K_m^I,
\]
where
\begin{equation}
K^I_m :=E^{Is}K_{sm},
\end{equation}
are the corresponding coordinates of the momentum space. In Eqs. (3.1) and (3.2) the vector fields $E^m_I$ are the triads on the spacelike hypersurface of spacetime, where the time coordinate $t$ is a constant. They are related to the metric tensor $q_{mn}$ induced on that hypersurface as:
\begin{equation}
E_{I}^mE_J^nq_{mn} = \delta_{IJ}
\end{equation}
for all $I, J = 1,2,3$.  Indeed, there is a canonical transformation from the phase space coordinates $(q_{mn}, p^{mn})$ of general relativity to the coordinates $(\tilde{E}^m_I, -\frac{1}{16\pi}K^I_m)$, provided that the resulting formulation of general relativity is assumed to remain invariant in the local SO(3) rotations of the triads. \cite{kymppi} Conversely, we may use the densitized triads $\tilde{E}_I^m$ as the coordinates of the momentum space, and the   quantities $\frac{1}{16\pi}K^I_m$ as the corresponding coordinates of the configuration space.

In the line element (2.15) we have:
\begin{subequations}
\begin{eqnarray}
q_{\lambda\lambda} &=& \frac{(y')^2}{4y},\\
q_{jk} &=& h_{jk},
\end{eqnarray}
\end{subequations}
for all $j, k = 1,2$. We may thus choose the triads such that
\begin{subequations}
\begin{eqnarray}
E^\lambda_3 &=& \frac{2\sqrt{y}}{y'},\\
E^j_IE^k_J h_{jk} &=&\delta_{IJ},
\end{eqnarray}
\end{subequations}
for all $I, J = 1,2$. So we find:
\begin{equation}
\tilde{E}^\lambda_3  = \sqrt{h},
\end{equation}
and we may write the Hamiltonian $H$ in Eq. (2.29) as:
\begin{equation}
H = \frac{a}{8\pi}\oint_B\tilde{E}^\lambda_3\,d^2\chi.
\end{equation}

   It is interesting that because the proper acceleration $a$ is assumed to be a constant, our Hamiltonian of the gravitational field may be expressed as a function of {\it just one} independent variable at each point of the spacelike two-surface $B$. Actually, this is something one might have expected: The Hamiltonian $H$ in Eq. (2.29) depends on the area $A$ of the two-surface $B$ only which, in turn, is determined by the metric tensor $h_{jk}$ induced on $B$. The metric tensor $h_{jk}$ has 3 independent components, but the requirement of the invariance of the gravitational action on the choice of the two coordinates $\chi^1$ and $\chi^2$ on $B$ brings along 2 constraints between these components. As a result, we are left with just $3-2=1$ independent, physical degree of freedom. 

  We now define the variables $p(\chi)$ associated with the points $\chi=(\chi^1,\chi^2)$ of the spacelike two-surface $B$ as:
\begin{equation}
p(\chi) := \frac{1}{8\pi}\tilde{E}_3^\lambda(\chi).
\end{equation}
By means of these variables we may write the area of $B$ as:
\begin{equation}
A = 8\pi\oint_B p(\chi)\,d^2\chi,
\end{equation}
and the Hamiltonian of the gravitational field takes the form:
\begin{equation}
H = a\oint_B p(\chi)\,d^2\chi.
\end{equation}
We shall now use the variables $p(\chi)$ as the coordinates of the momentum space. The corresponding coordinates of the configuration space we shall denote by $\phi(\chi)$. The quantity $\phi(\chi)$ associated with the point $\chi$ of $B$ obeys the Hamiltonian equation of motion:
\begin{equation}
\dot{\phi}(\chi) = \frac{\delta H}{\delta p(\chi)} = a,
\end{equation}
where the dot denotes the derivative with respect to the proper time $\tau$ of the observer. The general solution to Eq. (3.11) is:
\begin{equation}
\phi(\chi) = a\tau + C(\chi),
\end{equation}
where $C(\chi)$ is a constant function of the proper time $\tau$, and depends only on the point $\chi$ of $B$. Since the Hamiltonian $H$ in Eq. (3.10) does not involve $\phi(\chi)$ at all, the Hamiltonian equation of motion written for the momentum variable $p(\chi)$ reads as:
\begin{equation}
\dot{p}(\chi) = -\frac{\delta H}{\delta\phi(\chi)} = 0,
\end{equation}
and therefore the variables $p(\chi)$ are constants of motion of the system. 

  Now, what is $a\tau$? Eqs. (2.12) and (2.14) imply that during the lapse $dt$ of the Rindler time $t$ the elapsed proper time of the observer with constant proper acceleration $a$ is:
\begin{equation}
d\tau = \frac{1}{a}\,dt,
\end{equation}
and therefore:
\begin{equation}
dt = a\,d\tau.
\end{equation}
So we have, possibly up to an additive constant:
\begin{equation}
a\tau = t,
\end{equation}
and Eq. (3.12) takes the form:
\begin{equation}
\phi(\chi) = t + C(\chi).
\end{equation}
So we may identify the quantity $\phi(\chi)$, possibly up to an additive constant, as the {\it boost angle} of our observer at the point $\chi$. 

\section{Quantization}

When going over from the classical to the quantum-mechanical treatment of our model we replace the classical variables $\phi(\chi)$ and $p(\chi)$ by the corresponding operators $\hat{\phi}(\chi)$ and $\hat{p}(\chi)$ obeying the canonical commutation relations:
\begin{subequations}
\begin{eqnarray}
\lbrack\hat{\phi}(\chi), \hat{\phi}(\chi')\rbrack &=& \lbrack\hat{p}(\chi),\hat{p}(\chi')\rbrack = 0,\\
\lbrack\hat{\phi}(\chi),\hat{p}(\chi')\rbrack &=& i\delta(\chi-\chi'),
\end{eqnarray}
\end{subequations}
at all points $\chi, \chi'\in B$. In Eq. (4.1b) $\delta(\chi-\chi')$ is the delta function on the two-surface $B$. Eq. (3.10) implies that the Hamiltonian operator of the gravitational field takes the form:
\begin{equation}
\hat{H} = a\oint_B\hat{p}(\chi)\,d^2\chi,
\end{equation}
and hence the time-independent Schr\"odinger equation
\begin{equation}
\hat{H}\vert\psi\rangle = E\vert\psi\rangle
\end{equation}
written for the energy eigenstates $\vert\psi\rangle$ becomes to:
\begin{equation}
a\oint_B\hat{p}(\chi)\vert\psi\rangle\,d^2\chi = E\vert\psi\rangle.
\end{equation}
Since we may identify the energy $E$ with the classical Hamiltonian $H$ in Eq. (3.10), Eq. (4.4) reduces to:
\begin{equation}
\oint_B\hat{p}(\chi)\vert\psi\rangle\,d^2\chi = \oint_B p(\chi)\vert\psi\rangle\,d^2\chi.
\end{equation}
A solution to this equation is the one, where the state $\vert\psi\rangle$ is a common eigenstate of all of the operators $\hat{p}(\chi)$, {\it i. e.}
\begin{equation}
\hat{p}(\chi)\vert\psi\rangle = p(\chi)\vert\psi\rangle
\end{equation}
for all $\chi\in B$. In this equation $\hat{p}(\chi)$ is the operator associated with the point  $\chi$ and $p(\chi)$ the momentum variable at that point. We shall assume that the state $\vert\psi\rangle$ is normed to unity, {\it i. e.}
\begin{equation}
\langle\psi\vert\psi\rangle = 1.
\end{equation}
In this paper we shall consider a model of quantum gravity, where the energy eigenstates $\vert\psi\rangle$ satisfy both of the Eqs. (4.6) and (4.7), and see, where this will take us. 

In the {\it position representation} the quantum state $\vert\psi\rangle$ is described by a functional $\psi[\phi(\chi)]$ of the function $\phi(\chi)$. It may regarded as the wave function of the gravitational field. Between the states $\vert\psi_1\rangle$ and $\vert\psi_2\rangle$ represented by the wave functions $\psi_1[\phi(x)]$ and $\psi_2[\phi(\chi)]$, respectively, we define the inner product
\begin{equation}
\langle\psi_1\vert\psi_2\rangle := \int\psi^*_1[\phi(\chi)]\psi_2[\phi(\chi)]\,\mathcal{D}[\phi(\chi)],
\end{equation}
where $\mathcal{D}[\phi(\chi)]$ is an approriate integration measure in the space of the functions $\phi(\chi)$. In the position repesentation we write the operators $\hat{p}(\chi)$ as functional differential operators:
\begin{equation}
\hat{p}(\chi) = -i\frac{\delta}{\delta\phi(\chi)},
\end{equation}
and hence Eq. (4.6) takes the form:
\begin{equation}
-i\frac{\delta\psi[\phi(\chi)]}{\delta\phi(\chi)} = p(\chi)\psi[\phi(\chi)].
\end{equation}
The general solution to this equation is:
\begin{equation}
\psi[\phi(\chi)] = \mathcal{N}\exp\left\lbrack i\oint_B p(\chi)\phi(\chi)\,d^2\chi\right\rbrack,
\end{equation}
where $\mathcal{N}$ is an arbitrary complex number. Since the eigenstates associated with the different eigenvalues must be orthogonal, it follows that if we denote by $\psi[\phi(\chi)]$ the wave function associated with function $p(\chi)$, and by $\psi'[\phi(\chi)]$ the wave function associated with function $p'(\chi)$, then 
\begin{equation}
\int\psi^*[\phi(\chi)]\psi'[\phi(\chi)]\,\mathcal{D}[\phi(\chi)] = 1,
\end{equation}
when 
\begin{equation}
p(\chi) \equiv p'(\chi),
\end{equation}
and
\begin{equation}
\int\psi^*[\phi(\chi)]\psi'[\phi(\chi)]\,\mathcal{D}[\phi(\chi)] = 0,
\end{equation}
otherwise.

  $p(\chi)$ and $\phi(\chi)$ are functions of the coordinates $(\chi^1, \chi^2)$ of the points of a certain planar region $D$. We divide the planar region $D$ in $N$ non-intersecting subsets $D_j$ $(j = 1, 2,\dots, N)$ of $D$. In other words, $D_j\cap D_k=\emptyset$ for all $j\ne k$, and
\begin{equation}
D = D_1\cup D_2\cup\cdots\cup D_N.
\end{equation}
The area of the region $D$ is:
\begin{equation}
\mathcal{A} = \int_D d^2\chi.
\end{equation}
We shall assume that all of the subsets $D_j$ have the same area
\begin{equation}
\Delta\chi := \frac{\mathcal{A}}{N},
\end{equation}
and that the largest diameter of the subsets tends to zero, when $N$ tends to infinity. From each subset $D_j$ we pick up a point $\chi_j$. With these assumptions we may write the integral in the exponential in Eq. (4.11) as a Riemann integral:
\begin{equation}
\oint_B p(\chi)\phi(\chi)\,d^2\chi = \lim_{N\rightarrow\infty}\left(\sum_{j=1}^Np_j\phi_j\right)\,\Delta\chi,
\end{equation}
where we have denoted:
\begin{subequations}
\begin{eqnarray}
\phi_j &:=& \phi(\chi_j),\\
p_j &:=& p(\chi_j),
\end{eqnarray}
\end{subequations}
for all $j=1,2,\dots,N$. Hence we may write the wave functional $\psi[\phi(\chi)]$ as:
\begin{equation}
\psi[\phi(\chi)] = \mathcal{N}\lim_{N\rightarrow\infty}\prod_{j=1}^N[\exp(i p_j\phi_j\,\Delta\chi)].
\end{equation}

   Eq. (4.20) enables us to define the integral on the left hand side of Eq. (4.12):
 \begin{equation}
\int\psi^*[\phi(\chi)]\psi'[\phi(\chi)]\,\mathcal{D}[\phi(\chi)] := \mathcal{N}^*\mathcal{N}'\lim_{N\rightarrow\infty}\prod_{j=1}^N\left\lbrace\int_{-\frac{L}{2}}^{\frac{L}{2}}\exp\lbrack i(p'_j-p_j)\phi_j\,\Delta\chi\rbrack\,d\phi_j\right\rbrace.
\end{equation}
In this equation we have denoted $p'_j := p'(\chi_j)$ for all $j=1,2,\dots,N$. $L$ is a positive real and $\mathcal{N}'$ an arbitrary complex number. We could, of course, make $L$ to tend to infinity, which would mean that we integrate every variable $\phi_j$ from the negative to the positive infinity. Such an integration, however, would give infinite results, making impossible to satisfy the fundamental postulate (4.7) of our model. Because of that we shall assume from this point on that $L$ is a fixed, positive real, and see, where this will take us. In this sense $L$ is a free parameter of our model, and it must be fixed such that the predictions of the model are consistent with everything else we know about quantum gravity. 

   If we want the functions $\psi[\phi(\chi)]$ and $\psi'[\phi(\chi)]$ in Eq. (4.21) to satisfy the requirements (4.12) and (4.14), we must have:
\begin{equation}
p_j\,\Delta\chi = n_j\frac{2\pi}{L},
\end{equation}
where $n_j = 0, \pm 1, \pm 2,\dots$ for all $j= 1,2,\dots,N$. Eq. (3.9) implies that we may write the area of the spacelike two-surface $B$ as:
\begin{equation}
A = 8\pi\lim_{N\rightarrow\infty}\left(\sum_{j=1}^N p_j\,\Delta\chi\right),
\end{equation}
and so we may understand the quantity
\begin{equation}
A_j := 8\pi p_j\,\Delta\chi
\end{equation}
as the area of the piece, or constituent, of the two-surface $B$ associated with the subset $D_j$ of the planar region $D$. According to Eq. (4.22) the area eigenvalues of those constituents are of the form:
\begin{equation}
A_j = n_j\frac{16\pi^2}{L}.
\end{equation}
The concept of negative area obviously does not make sense, and therefore we shall abandon, from this point on, the negative values of the quantum numbers $n_j$, and keep only the non-negative ones. In other words, we shall assume that $n_j = 0,1,2,\dots$ for all $j=1,2,\dots,N$. 

   Eq. (4.25) implies that the area of an individual constituent of the surface $B$ is has a {\it discrete spectrum} with an equal spacing. Unless there is an infinite number of constituents with zero area on the surface $B$, we are therefore compelled to infer that the surface $B$ consists of a {\it finite  number} of constituents, each of them having an area spectrum given by Eq. (4.25). This number we shall denote by $N$. The total area of the surface is the sum of the areas of its constituents. In other words, we have:
\begin{equation}
A = A_1 + A_2 +\cdots + A_N,
\end{equation}
which means, through Eq. (4.24), that the area eigenvalues of the surface are of the form:
\begin{equation}
A = \frac{16\pi^2}{L}(n_1 + n_2 +\cdots + n_N).
\end{equation}
Hence we have shown that the fundamental postulates (4.6) and (4.7) of our model imply that the surface $B$, and thereby the horizon, has a {\it discrete structure}: The surface $B$ consists of a finite number of separate constituents, all of them having a discrete, equally spaced area spectrum. Consequently, the wave functional $\psi[\phi(\chi)]$ of spacetime in Eq. (4.11) becomes replaced by a function $\psi(\phi_1,\phi_2,\dots,\phi_N)$ of the $N$ boost angles $\phi_j$ $(j=1,2,\dots,N)$ at the constituents of the surface such that:
\begin{equation}
\psi(\phi_1,\phi_2,\dots,\phi_N) = \frac{1}{\sqrt{L^N}}\exp\left\lbrack\frac{i}{8\pi}(A_1\phi_1 + A_2\phi_2 +\cdots + A_N\phi_N)\right\rbrack.
\end{equation}
To satisfy the requirement (4.12) we have chosen the constant $\mathcal{N}$ to be $1/\sqrt{L^N}$. 

We have now completed the quantization of our model in the position representation, where  the wave function is expressed as a functional $\psi[\phi(\chi)]$ of the function $\phi(\chi)$. In the {\it momentum representation} the wave function is expressed as a functional $\tilde{\psi}[p(\chi)]$ of the function $p(\chi)$. When operating on $\tilde{\psi}[p(\chi)]$ by the operator $\hat{p}(\chi)$, we just multiply $\tilde{\psi}[p(\chi)]$ by $p(\chi)$. in other words,
\begin{equation}
\hat{p}(\chi)\tilde{\psi}[p(x)] = p(\chi)\tilde{\psi}[p(\chi)],
\end{equation}
no matter what is $\tilde{\psi}[p(\chi)]$. For instance, if we denote by $\tilde{\psi}_{p'}[p(\chi)]$ the eigenfunction associated with $p'(\chi)$, we must have:
\begin{equation}
\hat{p}(\chi)\tilde{\psi}_{p'}[p(\chi)] = p(\chi)\tilde{\psi}_{p'}[p(\chi)].
\end{equation}
However, since $\tilde{\psi}_{p'}[p(\chi)]$ is an eigenfunction associated with $p'(\chi)$, we must also have:
\begin{equation}
\hat{p}(\chi)\tilde{\psi}_{p'}[p(\chi)] = p'(\chi)\tilde{\psi}_{p'}[p(\chi)],
\end{equation}
and therefore the eigenfunction $\tilde{\psi}_{p'}[p(\chi)]$ must have the property:
\begin{equation}
p(\chi)\tilde{\psi}_{p'}[p(\chi)] = p'(\chi)\tilde{\psi}_{p'}[p(\chi)],
\end{equation}
no matter what is $p(\chi)$. 

  The only solution to this equation is the "delta functional" $\delta[p(\chi)-p'(\chi)]$, which has the property:
\begin{equation}
\delta[p(\chi)] = 0,
\end{equation}
whenever the function $p(\chi)$ is not identically zero. It also has the properties:
\begin{subequations}
\begin{eqnarray}
\int\delta[p(\chi)]\,\mathcal{D}[p(\chi)] &=& 1,\\
\int\delta[p(\chi)]F[p(\chi)]\,\mathcal{D}[p(\chi)] &=& F[0]
\end{eqnarray}
\end{subequations}
for an arbitrary functional $F[p(\chi)]$ of the function $p(\chi)$. $\mathcal{D}[p(\chi)]$ is an appropriate integration measure in the space of the functions $p(\chi)$. However, as we saw in Eq. (4.22), the quantities $p(\chi)$ may actually take just {\it discrete} values at the points of the surface $B$. Hence our delta functional $\delta[p(\chi)-p'(\chi)]$ is just an ordinary {\it Kronecker delta} with the properties:
\begin{equation}
\delta[p(\chi)-p'(\chi)] = 1,
\end{equation}
whenever $p(\chi)\equiv p'(\chi)$, and
\begin{equation}
\delta[p(\chi)-p'(\chi)] = 0,
\end{equation}
otherwise.

   As the reader may have noticed, quantization of gravity in our  model from the point of view of an observer with constant proper acceleration just outside of a horizon of spacetime is really very similar to the quantization of a system of {\it free particles} in non-relativistic quantum mechanics: As in the system of free particles, the coordinates of the momentum space were constants of motion, whereas the corresponding coordinates of the configuration space were, up to additive constants, linear functions of the time coordinate of the observer. The constant $L$ introduced in this section may be understood as the edge length of a box used in the {\it box normalization} of the wave function in the position representation. However, whereas in the box normalization in non-relativistic quantum mechanics the edge length $L$ is finally taken to infinity at the end of the calculations, in our model $L$ is assumed to be a {\it fixed number}.

\section{Thermodynamical Properties of Spacetimes with Horizon}

Our simple model about the quantum-mechanical properties of the horizons of spacetime implied that the horizon of spacetime consists of separate constituents, each of them contributing to the horizon an area, which is an integer times a certain fundamental area. In the SI units the horizon area takes the form:
\begin{equation}
A = \alpha\ell_{Pl}^2(n_1 + n_2 + \cdots + n_N),
\end{equation}
where $N$ is the number of the constituents of the horizon, and
\begin{equation}
\ell_{Pl} := \sqrt{\frac{\hbar G}{c^3}}\approx 1.6\times 10^{-35}m
\end{equation}
is the Planck length. We have denoted: 
\begin{equation}
\alpha:=\frac{16\pi^2}{L}.
\end{equation}
The quantum numbers $n_1, n_2,\dots, n_N$ are non-negative integers, and they determine the {\it area eigenstates} of the constituents of the horizon. Constituent $j = 1, 2,\dots, N$ is in {\it vacuum}, if $n_j = 0$; otherwise the constituent in an {\it excited state}. 

  It is interesting that this very simple microscopic model of the horizons implies, among other things, all of the well-known results, such as the Hawking effect, for the thermodynamics of black holes. The only necessary assumption needed for the derivation of these results is that the microscopic states of the horizon are identified with the combinations of the excited states of its constituents such that when the excited states of the constituents, determined by the quantum numbers $n_j$, are permuted, the microscopic state of the horizon will also change. That derivation has been carried out in various ways and in various contexts in several papers. $[11-16]$ For the benefit of the reader, we shall now outline its main points.

    As we found in Section 2, beginning from the standard ADM formulation of general relativity, the quantity
\begin{equation}
H = \frac{a}{8\pi}A
\end{equation}
may be viewed as the classical Hamiltonian, and hence energy of the gravitational field, inside of a closed spacelike two-surface with area $A$, just outside of the horizon, from the point of view of an observer lying on the surface with constant proper acceleration $a$. Since the 
two-surface is assumed to lie just outside of the horizon, we may identify its area $A$ with the area of the horizon. 

  The thermodynamical properties of any system may be obtained from its partition function
\begin{equation}
Z(\beta) :=\sum_n e^{-\beta E_n},
\end{equation}
where $\beta$ is the temperature parameter, and we have summed over all energy eigenstates $n$ with energy eigenvalues $E_n$ of the system. Using Eqs. (5.1) and (5.4) and employing the assumptions stated above we find that the partition function of the horizon takes, from the point of view of our observer, the form: $[11,14,16]$
\begin{equation}
\begin{split}
Z(\beta) = &\sum_{n_1=1}^\infty\exp\left(-n_1\frac{\alpha\beta a}{8\pi}\right)\\
                +&\left\lbrack\sum_{n_1=1}^\infty\exp\left(-n_1\frac{\alpha\beta a}{8\pi}\right)\right\rbrack\left\lbrack\sum_{n_2=1}^\infty\exp\left(-n_2\frac{\alpha\beta a}{8\pi}\right)\right\rbrack\\
                + &\cdots\\
                +&\left\lbrack\sum_{n_1=1}^\infty\exp\left(-n_1\frac{\alpha\beta a}{8\pi}\right)\right\rbrack\cdots\left\lbrack\sum_{n_N=1}^\infty\exp\left(-n_N\frac{\alpha\beta a}{8\pi}\right)\right\rbrack.
\end{split}
\end{equation}
In the first term on the right hand side of Eq. (5.6) just one of the constituents of the horizon is in an excited state. In the second term two of the constituents are in excited states. Finally, in the last term all $N$ of the constituents of the horizon are in excited states. 

    The partition function written in Eq. (5.6) may be evaluated explicitly. A straightforward 
 calculation, based on the elementary properties of the geometric series, gives the result: \cite{kuutoo}
\begin{equation}
Z(\beta) = \frac{1}{2 - 2^{\beta T_C}}\left\lbrack 1 - \left(\frac{1}{2^{\beta T_C} - 1}\right)^N\right],
\end{equation}
where we have defined the {\it characteristic temperature}
\begin{equation}
T_C := \frac{\alpha a}{8\pi\ln (2)}
\end{equation}
of the horizon. Eq. (5.7) is valid, whenever $\beta\ne \frac{1}{T_C}$. If $\beta = \frac{1}{T_C}$, we have $Z(\beta) = N$. From this partition function one may obtain, for instance, an expression for the energy of the gravitational field per a constituent of the horizon:
\begin{equation}
\begin{split}
\bar{E} &:=-\frac{1}{N}\frac{\partial}{\partial\beta}\ln\left\lbrack Z(\beta)\right\rbrack\\
           &= \left\lbrack\frac{1}{N}\frac{2^{\beta T_C}}{2^{\beta T_C} - 2} + \frac{2^{\beta T_C}}{2^{\beta T_C} - 1 - (2^{\beta T_C} - 1)^{N+1}}\right\rbrack T_C\ln(2).
\end{split}
\end{equation}

   In all of our calculations, which so far have been carried explicitly, with no approximations made, we have always assumed that the number $N$ of the constituents of the horizon is very large. With this assumption we find that the first term on the right hand side of Eq. (5.9) will always vanish, when $\beta T_C\ne 1$. The second term in the brackets will also vanish in the large $N$ limit, whenever $\beta T_C >1$. This means that the constituents of the horizon are effectively in the vacuum, when the temperature of the horizon is, from the point of view of our observer, less than the characteristic temperature $T_C$. However, something very strange happens to $\bar{E}$, when $\beta T_C< 1$, which means that the temperature exceeds the characteristic temperature $T_C$. When $\beta T_C <1$, the quantity $(2^{\beta T_C} - 1)^{N+1}$ vanishes in the large $N$ limit, and we may write the average energy $\bar{E}$ in Eq. (5.9), in  effect, as:
\begin{equation}
\bar{E} = \frac{2^{\beta T_C}}{2^{\beta T_C} - 1}T_C\ln(2).
\end{equation}
So we find that when $\beta T_C$ tends to 1 from the left hand side, which means that the temperature $T$ of the horizon tends to $T_C$ from the right hand side, we have:
\begin{equation}
\bar{E} = 2T_C\ln(2) = \frac{2\alpha a}{8\pi},
\end{equation}
where we have used Eq. (5.8). This means that at the characteristic temperature $T_C$ the constituents of the horizon jump, in average, from the vacuum to the second excited states.  In other words, the horizon performs a {\it phase transition} at the characteristic temperature $T_C$. Since the consituents are effectively in the vacuum, when $T<T_C$, and jump to the second excited states, when $T=T_C$, we may view the characteristic temperature $T_C$, defined in Eq. (5.8), as the {\it lowest possible temperature} of the horizon.

  As we found in Eq. (5.8), the characteristic temperature $T_C$ depends on the parameter $\alpha$. It is interesting to observe that if we put:
\begin{equation}
\alpha = 4\ln(2),
\end{equation}
then
\begin{equation}
T_C = \frac{a}{2\pi},
\end{equation}
which agrees with the {\it Unruh temperature} \cite{seetoo}
\begin{equation}
T_U := \frac{a}{2\pi}
\end{equation}
measured by an observer with proper acceleration $a$. To get the temperature of the horizon from the point of view of a distant observer we must replace the proper acceleration $a$ by the {\it surface gravity} $\kappa$ at the horizon. The surface gravity $\kappa$ tells the proper acceleration at the horizon as seen by the distant observer. If the horizon under consideration happens to be a black hole event horizon, we therefore find that the lowest  possible temperature of the black hole is, from the point of view of our observer:
\begin{equation}
T_\infty = \frac{\kappa}{2\pi}.
\end{equation}
This is one of the most fundamental results of black-hole thermodynamics. \cite{kasitoo}

   Eq. (5.13) has important implications to black hole entropy. As it is well known, between the temperature parameter $\beta$, entropy $S$ and energy $E$ of any system there is the relationship:
\begin{equation}
\beta = \frac{\partial S}{\partial E}.
\end{equation}
Identifying the right hand side of Eq. (5.4) as the energy $E$ of a black hole from the point of view of an observer with proper acceleration $a$, just outside of its event horizon, and putting $\beta = \frac{1}{T_C}$, we find:
\begin{equation}
\frac{2\pi}{a} = \frac{\partial S}{\partial A}\left(\frac{dE}{dA}\right)^{-1} = \frac{\partial S}{\partial A}\frac{8\pi}{a},
\end{equation}
which implies:
\begin{equation}
S = \frac{1}{4}A.
\end{equation}
In other words, the black hole entropy is, in the natural units, exactly one-quarter of its event horizon area, when its temperature agrees with $T_C$. \cite{ysitoo} The same result may be obtained by means of a systematic analysis, beginning from the partition function $Z(\beta)$ obtained in Eq. (5.7). \cite{yytoo}

    As we have seen, our extremely simple model predicts all of the classic properties of  black-hole thermodynamics: With the choice (5.12) for the parameter $\alpha$ we find that  the black hole has temperature $\frac{\kappa}{2\pi}$, and its entropy is one-quarter of its event horizon area. Actually, the model does even more. There is a well-known problem concerning the Schwarzschild black hole in a heat bath: When the hole absorbs radiation from its environment, its Schwarzschild mass $M$ increases. However, Eq. (5.15) implies that the temperature of the hole is
\begin{equation}
T_\infty = \frac{1}{8\pi M},
\end{equation}
which means that the temperature of the hole {\it decreases}, when its mass $M$ increases, and therefore the hole absorbs heat faster and faster. One may show that the mass $M$ of the Schwarzschild black hole becomes infinite in a {\it finite time}. This is known as the {\it singularity problem} in the black hole mass. \cite{kakskyt}

  Our model solves the singularity problem in a very nice way: If we look at Eq. (5.9), we observe that the average energy $\bar{E}$ per a constituent depends on the temperature parameter $\beta$, and thus on the temperature $T$. When the temperature $T$ is increased above $T_C$, the constituents begin to jump to higher excited states, and the mass $M$ of the hole increases. Eq. (5.10) implies that the relationship between the temperature $T_\infty$ measured by a distant observer and the mass $M$ of the hole is, in the large $N$ limit, not given by Eq. (5.19), but by an equation: \cite{kuutoo}
\begin{equation}
T_\infty = -\frac{1}{8\pi M}\frac{\ln(2)}{\ln\left[1 - \frac{N\ln(2)}{4\pi M^2}\right]},
\end{equation}
which is an {\it increasing} function of $M$. Hence it follows that when the black hole absorbs heat and its mass increases, the temperature of the hole does not really decrease, but it increases. During the process, where the hole tends to a thermal equilibrium with the heat bath, its temperature increases, and it absorbs heat all the time slower and slower. A study reveals that the mass of the hole tends to a certain {\it finite} value, which depends on the temperature of the heat bath. \cite{kuutoo} So, there is no singularity in the black hole mass. 

\section{Consequences}

As we have seen, our simple model of quantum gravity predicts, starting from the basic principles, the horizon area spectrum in Eq. (5.1). Hence it predicts, among other things, the well-known results of black-hole thermodynamics, and it also provides a solution to the problem of the mass singularity of black holes. When expressed in terms of the areas $A_j$ of the constituents of the horizon the wave function of the gravitational field takes a simple and beautiful form expressed in Eq. (4.28).

       There is, however, a somewhat annoying feature in our model of quantum gravity: It involves a new, dimensionless parameter $L$, which we introduced by hand as the edge length of an $N$-dimensional box used in the box normalization of the wave function of the gravitational field. The edge length $L$ is related to the parameter $\alpha$ through Eq. (5.3), and if we use Eq. (5.12), we find:
\begin{equation}
L = \frac{4\pi^2}{\ln(2)}\approx 57.0.
\end{equation}
  Our parameter is not any obvious combination of the natural constants, and we did not take $L$ to infinity at the end of the calculations, either, as we do, when we box normalize the wave function of the system of free particles in non-relativistic quantum mechanics. Instead, $L$ is a {\it fixed number}, whose value is dictated by the need of our model to predict the standard results of black-hole thermodynamics. In this sense the parameter $L$ plays in our model of quantum gravity the same role as does the Barbero-Immirzi parameter $\gamma$ in loop quantum gravity. \cite{kakskyta}

     Curiously, these disadvantages of our model may actually be turned to its advantage, but they take us to an analysis of motion in quantized spacetime. The boost angle $\phi$ determines the velocity $v$ of an observer through the relation:
\begin{equation}
v = \tanh(\phi).
\end{equation}
It is natural to use $L$ as a bound for the allowed values of $\phi$ such that
\begin{equation}
-\frac{L}{2}\le \phi\le\frac{L}{2}.
\end{equation}
In this way the points $(\phi_1,\phi_2,\dots, \phi_N)$ of the configuration space are really confined to lie in an $N$-dimesional box with edge length $L$. We chose the allowed values of $\phi$ as in Eq. (6.3), because there is no reason to favor any direction, when the observer moves in spacetime. We could, of course, have chosen the allowed values of $\phi$ to lie, for instance, within the interval $[0,L]$, but in that case any possibility of motion to the negative direction would have been excluded, which is absurd. For the same reason we did not perform the half-line quantization, where the quantities $\phi_j$ would have been permitted to take positive values only, to our system.

   Eq. (6.3) sets, through Eqs. (6.1) and (6.2), the following bounds to the velocity $v$ of the observer:
\begin{equation}
-c\tanh\left\lbrack\frac{2\pi^2}{\ln(2)}\right\rbrack \le v\le c\tanh\left\lbrack\frac{2\pi^2}{\ln(2)}\right\rbrack.
\end{equation} 
Actually, these bounds are pretty wide. Because
\begin{equation}
\tanh(\phi) = \sqrt{1 - \frac{1}{\cosh^2(\phi)}},
\end{equation}
for positive $\phi$, we find that the maximum speed of the observer is:
\begin{equation}
\begin{split}
v_{max} &= c\sqrt{1 - \frac{1}{\cosh^2\left\lbrack\frac{2\pi^2}{\ln(2)}\right\rbrack}}\\
              &\approx\left\lbrace 1 - \frac{1}{2\cosh^2\left\lbrack\frac{2\pi^2}{\ln(2)}\right\rbrack}\right\rbrace c\\
             &\approx (1 - 3.7\times 10^{-25})c.
\end{split}
\end{equation}
Hence we note that the observer may still attain velocities, which are very close to the speed $c$ of light. If a material particle raced with a photon through a galaxy with diameter $10^5$ light years with this speed, it would lag less than 0.4 mm behind the photon at the end of the race. 

   In addition of giving the maximum speed of the observer, Eq. (6.3) also gives the maximum speed of massive particles with respect to the observer. To see that this is the case, we first note that its always possible to pick up, at the given point of spacetime, an inertial frame of reference for the observer. In this frame the observer has a certain proper acceleration $a$, and a certain boost angle $\phi$. The proper acceleration $a$ of the observer is always the same, no matter how we choose the frame, but the boost angle $\phi$ is not, and it depends on the choice of the frame, which may be done arbitrarily. It is always possible to pick up for a massive particle an inertial frame of reference, where the particle is at rest. We may take that frame as the inertial frame of reference of the observer. According to our results it is impossible to accelerate our observer to a speed exceeding $v_{max}$ in this frame. Hence the speed of the particle with respect to the observer will not exceed $v_{max}$, either.

  Since the energy $E$ of the particle with mass $m$ is related to its boost angle $\phi$ as:
\begin{equation}
E = mc^2\cosh(\phi),
\end{equation}
we find that the maximum energy of the particle is:
\begin{equation}
E_{max} = mc^2\cosh\left\lbrack\frac{2\pi^2}{\ln(2)}\right\rbrack.
\end{equation}
For the electron, for instance, the maximum energy is
\begin{equation}
E_{max}^{electron} \approx 6.0\times^{17}eV,
\end{equation}
whereas for the proton the maximum energy is
\begin{equation}
E_{max}^{proton}\approx 1.1\times 10^{21}eV.
\end{equation}
Hence our model makes a concrete and precise observational prediction: It is impossible for an electron to reach energies higher than $6.0\times 10^{17}eV$, and for a proton energies higher than $1.1\times 10^{21}eV$. Should there ever be an observation of an electron with energy higher than $6.0\times 10^{17}eV$, or of a proton with energy higher than $1.1\times 10^{21}eV$, our model of quantum gravity would have been proved wrong.  It is interesting that both of these energies are well below the Planck energy, which is
\begin{equation}
E_{Pl} := \sqrt{\frac{\hbar c^5}{G}}\approx 1.2\times 10^{28}eV.
\end{equation}
Our prediction may therefore be viewed as an entirely new, unexpected effect of quantum gravity. 

      The most highly energetic particles have been found in cosmic rays. The particles with extremely high energies are mostly protons. Now and then protons with energies of the order of $10^{19}eV$ or higher are observed. \cite{yy} The highest energy ever measured for a particle in cosmic radiation has been $(3.2\pm 0.9)\times 10^{20}eV$. \cite{kaa} It is not known, what this the so-called "Oh-My-God particle" was, but it is believed to have been a proton. So we find that the very highest energies ever measured for the protons in cosmic rays are roughly of the same order of magnitude as in the prediction made in Eq. (6.10) for their highest possible energies. The result may be viewed as an observational support for our model of quantum gravity. 

No doubt, even the particles with energies close to the maximum energy $E_{max}$ given by Eq. (6.8) will undergo a red shift, if they arrive from long enough distances. Employing  for the red shift $z$ the standard formula
\begin{equation}
z = \frac{H_0D}{c},
\end{equation}
where $H_0$ is the Hubble constant,  one finds that the red-shifted maximum energy of a particle arriving from distance $D$ is
\begin{equation}
E'_{max} = \frac{E_{max}}{\frac{H_0D}{c}+1}.
\end{equation}
For instance, if $D$ equals, say, 3 billion light years, or $3\times 10^{25}m$, and we put $2.2\times 10^{-18}s^{-1}$ for $H_0$, one obtains for the red-shifted maximum energy an approximation
\begin{equation}
E'_{max}\approx 0.8E_{max}.
\end{equation}

\section{Concluding Remarks}

In this paper we have constructed, starting from the first principles and the ADM formulation of general relativity, a simple model of quantum gravity from the point of view of an observer with constant proper acceleration $a$, just outside of a horizon of spacetime. Curiously, quantization of gravity according to our model was very similar to the quantization of a system of free particles in non-relativistic quantum mechanics. In the process we were compelled to introduce a sort of box normalization to the quantum states of the gravitational field. The edge length $L$ of the box then appeared as a parameter of our model, and when $L$ was chosen appropriately, the model re-produced all of the well-known results of black-hole thermodynamics. Actually, the model did even more, since it provided a solution to the so-called mass singularity problem of the Schwarzschild black hole.

   The most startling result of our model, however, was that it predicted a certain upper limit to the speed of material particles, which is a very tiny amount {\it less} than the speed of light, depending on the choice for the parameter $L$ of our model. This result was an outcome of the requirement, posed as a postulate of our model, that the energy eigenstates of the gravitational field must be normalizable. We found that to satisfy this requirement the horizon must consist of a finite number of separate pieces, or constituents, each of them having a discrete area spectrum with an equal spacing. The wave function of the gravitational field was expressed in terms of the boost angles associated with these constituents, and the normalizability of the wave function put certain restrictions to the boost angle, and hence the velocity, of the observer.  Since velocity is relative, the restrictions to the velocity of the observer may be viewed as restrictions to the velocity of the material particles with respect to the observer. 

       The restrictions for the velocities of the material particles bring along restrictions to the possible energies of the particles. In other words, the energy of a material particle has a certain absolute upper bound. For protons, for example, this upper bound was found to be $1.1\times 10^{21}eV$. The result is interesting, because the very highest energies ever measured for protons in cosmic rays have been around $10^{19} eV$, some of them being even $10^{20}eV$. This is roughly of the same order of magnitude as is the highest possible energy of the proton according to our model, still being very much less than the Planck energy, which is around $10^{28}eV$. 

  It must be emphasized once more again that our results have been obtained by means of a straightforward application of the standard rules of quantum mecanics to Einstein's general relativity. No doubt, our tentative model of quantum gravity still has several shortcomings, such as our decision to exclude the negative area eigenvalues "by hand".   Nevertheless, our model certainly has an advantage that it really makes a concrete and precise prediction, whose validity may be tested by means of currently available methods.

\bigskip

{\bf Data Availability Statement}

\medskip

The relevant data needed for the understanding of this paper is all included either in the references listed in below, or in the calculations and the analysis worked out in the text.

\end{document}